# Wind-Wave Model with an Optimized Source Function[1]

**V. G. Polnikov**

*Oboukhov Institute of Atmospheric Physics, Russian Academy of Sciences,
Pyzhevskii per. 3, Moscow, 119017 Russia*
*e-mail: polnikov@mail.ru*

**Abstract**—On the basis of the author's earlier results, a new source function for a numerical wind-wave model optimized by the criterion of accuracy and speed of calculation is substantiated. The proposed source function includes (a) an optimized version of the discrete interaction approximation for parametrization of the nonlinear evolution mechanism, (b) a generalized empirical form of the input term modified by adding a special block of the dynamic boundary layer of the atmosphere, and (c) a dissipation term quadratic in the wave spectrum. Particular attention is given to a theoretical substantiation of the least investigated dissipation term. The advantages of the proposed source function are discussed by its comparison to the analogues used in the widespread models of the third generation WAM and WAVEWATCH. At the initial stage of assessing the merits of the proposed model, the results of its testing by the system of academic tests are presented. In the course of testing, some principals of this procedure are formulated. The possibility of using the testing results to study the physics of evolution processes in wind waves is shown. It is noted that the specially added block of the dynamic boundary layer of the atmosphere makes it possible to give an exhaustive description of the air–sea-interface's characteristics, which may be used to improve wave forecasting. This new modeling quality allows us to make a statement about the construction of a model of the next (fourth) generation.

## 1. INTRODUCTION

It has long been found that the problem of wind-wave numerical modeling and forecasting is important from both scientific and practical points of view [1, 2]. For this reason, a great deal of investigations in this field have been formed as a separate scientific topic. This is testified by an imposing list of famous scientists engaged in this topic (for example, see references in [1, 2]) and by a series of international projects, such as SWAMP [3], WAMDI [4], WISE [5], etc. The most brilliant results of these projects are the models of the third generation WAM [4], WAVEWATCH (WW) [6], and SWAN [5], which are widely used abroad. These models have not found wide use in Russia for a number of reasons. On the one hand, along with doubtless merits, these models have noticeable drawbacks, both in respect of physical substantiation and in the technique of implementing numerical procedures [7]. On the other hand, the numerical models of earlier generations available in Russia work fairly well in coping with practical problems of a certain level [8]. This factor restrain wide use of the foreign and domestic models of next generations in Russia [9, 10]. However, owing to arising new fields of wave-model applications (for example, calculations of boundary-layer characteristics [11] or upper-ocean parameters [12]), the necessity of promoting more sophisticated models becomes more and more evident.

At the same time, to date, various authors have derived important theoretical results making it possible to improve the third-generation wind-wave models mentioned above and formulate an approach to the construction of models of the fourth generation. It has been proposed to take all the best existing models and to supplement the model by new theoretically substantiated parametrizations, thus giving it a new quality. As the basis for such a model improvement, a special criterion of accuracy and speed of wave calculation is used. In some cases, this criterion is formulated in an explicit form (for example, in paper [13], devoted to the optimization of a parametrization of nonlinear evolution mechanisms for wind waves). In a general case, this criterion is speculative. As a rule, it is based on an explicit analytic comparison of advantages and disadvantages between the previous and new parametrical representations of evolution mechanisms for wind waves. In a scientific aspect, the formulation of the problem under consideration is the following.

The evolution equation for wind waves is usually written in the form of a transport equation for the two-dimensional energy spectrum of the surface elevation given in the frequency–angle representation $S \equiv S(\sigma, \theta, \mathbf{x}, t)$ [1, 2]. Here, $\sigma$ and $\theta$ are the frequency and the angle of a wave spectral component, respectively; $\mathbf{x} = (x, y)$ is the vector of the horizontal coordinates; and $t$ is the time variable. In such a case, for deep water with no currents, the spa-

---

[1] The article was translated by the author.

tiotemporal evolution equation for wind waves has the form

$$\frac{\partial S}{\partial t} + C_{gx}\frac{\partial S}{\partial t} + C_{gy}\frac{\partial S}{\partial y} = F \equiv \text{NL} + \text{IN} - \text{DIS}. \quad (1)$$

Here, $(C_{gx}, C_{gy})$ is the group-velocity vector corresponding to the proper wave component, which is defined as

$$\mathbf{C}_g = \frac{\partial \sigma(k)}{\partial k}\frac{\mathbf{k}}{k} = (C_{gx}, C_{gy}), \quad (2)$$

while the dependence of the frequency $\sigma(k)$ on the wave vector $\mathbf{k}$ is given by the dispersion relation for deep-water waves

$$\sigma = \sqrt{gk}. \quad (3)$$

The left-hand side of Eq. (1) corresponds to the advection part of the model, which is not discussed here. The physical content of the model is in the source function $F$, including three terms—three parts of the evolution mechanism for wind waves:

the rate of nonlinear energy transfer through a wave spectrum, Nl (nonlinearity term);

the rate of energy transfer from wind to waves, In (input term); and the rate of wave energy loss through wave interactions with the upper-layer turbulence, Dis (dissipation term).

It is a specific form of the mathematical expressions (parametrizations) used for the terms of the source function that determines a physical specification of each model. As was mentioned above, the most famous prototypes of wind-wave models reasonable for comparison are the WAM [4] and WW [6] models.

In this paper, a critical analysis will be carried out for parametrizations of individual terms of the source function for the models mentioned above and more sophisticated versions for evolution mechanisms will be proposed that are based on modern results obtained in this field. Special attention will be given to a theoretical substantiation of the least investigated mechanism of wave-energy dissipation. A standard approach to describing the input mechanism will be added by the block of dynamic boundary layer, which is based on the recent paper by Makin and Kudryavtsev [14]. At the same time, certain simplifications will be proposed for the existing complicated parametrizations to save computer time and to enhance the speed of wave calculations without loss of accuracy (the criterion of accuracy and speed). This is the main aim of this study.

The properties of the new model will be demonstrated by the results of model testing by the available tests that have shown their high informativeness [1]. Additionally, the necessity of using a proper physical model similar to that proposed in [14] to calculate real characteristics for the atmospheric boundary layer will be demonstrated.

The paper is organized as follows. In Sections 2–4, new optimized parametrizations are described for the nonlinear term of the source function Nl, for the input term In, and for the dissipation term Dis, respectively. In Section 5, a short description of general principles of testing is given. In Section 6, the results of testing are presented and their analysis is given from the standpoint of the quality of describing the physics of wave-evolution processes. Conclusions are presented in Section 7.

## 2. PARAMETRIZATION OF THE NONLINEAR EVOLUTION MECHANISM (NL TERM)

This mechanism is fundamentally important for description of wave development. In the context of certain approximations, the nonlinear mechanism of evolution has been studied almost exhaustively, starting with the pioneering paper by Hasselmann [15], which was elaborated later in numerous papers by Zakharov [16], and ending with different numerical parametrizations of the Nl term (a representative list of references can be found, for example, in [1, 2, 13]). According to the weak turbulence theory (according to Zakharov's terminology), in the case of the wave action spectrum representation, the $N(\mathbf{k})$-term Nl[$N(\mathbf{k})$] is described by the four-wave kinetic equation of the form [15, 16]

$$\begin{aligned}\text{Nl}[N(\mathbf{k}_4)] = 4\pi \int d\mathbf{k}_1 \int d\mathbf{k}_2 \int d\mathbf{k}_3 M^2(\mathbf{k}_1, \mathbf{k}_2, \mathbf{k}_3, \mathbf{k}_4) \\
\times [N(\mathbf{k}_1)N(\mathbf{k}_2)(N(\mathbf{k}_3) + N(\mathbf{k}_4)) \\
- N(\mathbf{k}_3)N(\mathbf{k}_4)(N(\mathbf{k}_1) + N(\mathbf{k}_2))] \\
\times \delta(\sigma(k_1) + \sigma(k_2) - \sigma(k_3) - \sigma(k_4)) \\
\times \delta(\mathbf{k}_1 + \mathbf{k}_2 - \mathbf{k}_3 - \mathbf{k}_4).\end{aligned} \quad (4)$$

Here, $\mathbf{k}_i$ is the wave vector corresponding to the wave component with frequency–angle parameters $(\sigma_i, \theta_i)$ ($i = 1, 2, 3, 4$); $M(\ldots)$ is the matrix elements of four interacting waves, and $\delta(\ldots)$ is the Dirac delta function testifying to a resonance feature of nonlinear interactions. The transition from the wave-action spectrum $N(\mathbf{k})$ to the wave energy spectrum $S(\sigma, \theta)$ is carried out with the relationship

$$S(\sigma, \theta) = \frac{\sigma}{4\pi^2 g}N(\sigma, \theta) = \frac{\sigma}{4\pi^2 g}\frac{k}{C_g}N(\mathbf{k}), \quad (5)$$

where $g$ is the gravity acceleration and $\mathbf{C}_g$ is given by the relation (2).

Owing to the complexity of calculation of the kinetic integral on the right-hand side of (4), in a numerical model, it is necessary to use a certain approximation. In [13], it was shown that, in accordance with the accuracy–speed criterion (for the definition of criterion, see [13]), the best approximation, among all the known approximations substantiated theoretically, is the discrete-interaction approximation (DIA). According to [17], the essence of the DIA lies in the following. Instead of the total kinetic integral in (4), the DIA approximation uses only one term with a specially chosen configuration of the four interacting wave vectors $\mathbf{k}_1, \mathbf{k}_2, \mathbf{k}_3, \mathbf{k}_4$ (see Fig. 1) that meet the following four-wave resonance conditions:

$$\mathbf{k}_1 + \mathbf{k}_2 = \mathbf{k}_3 + \mathbf{k}_4, \qquad (6a)$$

$$\sigma_1 + \sigma_2 + \sigma_3 + \sigma_4. \qquad (6b)$$

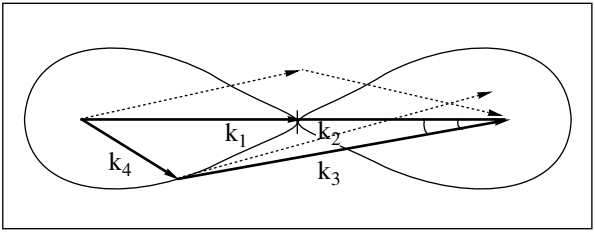

**Fig. 1.** Scheme of choice for the four interacting vectors in the original version of the DIA (solid vectors) and in the version of the FDIA (dotted vectors).

In such a case, the term for the nonlinear evolution mechanism assumes the simplest form among all possible forms. Since the values of the nonlinear term function $Nl[S(\sigma, \theta)]$ for Eq. (1) in a spectral representation are needed at each point of the numerical frequency–angle grid $\{\sigma, \theta\}$, the loop is made for the array of grid points. In the loop mentioned, the estimate of the term $Nl[S(\sigma, \theta)] \equiv NL(\sigma, \theta)$ is determined by the following relations [17]:

$$Nl(\sigma, \theta) = I(\sigma_1, \theta_1, \sigma_2, \theta_2, \sigma_3, \theta_3, \sigma, \theta), \qquad (7a)$$

$$Nl(\sigma_3, \theta_3) = I(\sigma_1, \theta_1, \sigma_2, \theta_2, \sigma_3, \theta_3, \sigma, \theta), \qquad (7b)$$

$$Nl(\sigma_1, \theta_1) = -I(\sigma_1, \theta_1, \sigma_2, \theta_2, \sigma_3, \theta_3, \sigma, \theta), \qquad (7c)$$

$$Nl(\sigma_2, \theta_2) = -I(\sigma_1, \theta_1, \sigma_2, \theta_2, \sigma_3, \theta_3, \sigma, \theta), \qquad (7d)$$

where

$$I(\sigma_1, \theta_1, \sigma_2, \theta_2, \sigma_3, \theta_3, \sigma, \theta)$$
$$= C\sigma^{11}[S_1 S_2 (S_3 + (\sigma_3/\sigma)^4 S) \qquad (8)$$
$$- S_3 S((\sigma_2/\sigma)^4 S_1 + (\sigma_1/\sigma)^4 S_2)].$$

In formulas (7) and (8), the index 4 is omitted, $S_i = S(\sigma_i, \theta_i)$, and the values of $(\sigma_i, \theta_i)$ for the indices $i = 1, 2, 3$ are given by certain relations following from a specific configuration of the four interacting vectors (for details, see [13, 17]). In the original version of the DIA, the fitting constant $C$ in (8) is taken equal to $3 \times 10^7$ [17]. This version of the DIA is used in the WAM and WW models with some variations in the constant $C$.

The principal (technical) disadvantage of the original version of the DIA is the necessity to carry out an interpolation of spectrum values $S_i = S(\sigma_i, \theta_i)$ for those points $(\sigma_i, \theta_i)$ that deviate from the points of the numerical grid used for calculations. This factor substantially reduces the speed of estimation of the Nl term.

In application aspects, the most important features of the DIA are as follows [13]:

(i) In the WAM model, the relative amount of time needed to calculate the term Nl by the DIA is about 48% of the whole time of wave forecasting;

(ii) In the DIA approximation, the relative error averaged over a representative series of two-dimensional spectral shapes has an order of 60%.

A full study of the DIA, its comparison with other approximations, and a modification of the DIA were carried out in [13, 18]. In [13], an accelerated version of the DIA (called the fast DIA (FDIA)) was proposed. In [18], a new configuration of the four interacting vectors was found, which is more effective with the criterion of accuracy and speed of calculations. Both of these results allow an essential optimization of the calculation of the term $Nl(\sigma, \theta)$.

The essence of the accelerated version of the DIA (FDIA) consists in discarding the exact-resonance conditions (6a) and (6b) by changing the configuration of four interacting vectors $\mathbf{k}_1, \mathbf{k}_2, \mathbf{k}_3, \mathbf{k}_4$ in such a manner that each of the vectors is located at a node of the numerical grid $\{\sigma, \theta\}$ (see Fig. 1). In this case, the necessity of interpolating the spectrum for calculation of (8) is eliminated, thereby markedly accelerating the procedure of calculating the term $Nl(\sigma, \theta)$. As was shown in [13], in the case of a sufficiently fine numerical grid, the proposed changes in the configuration enhance even the accuracy somewhat rather than lead to its loss.

An additional variation of parameters for the configuration used in FDIA, which was carried out in [18], made it possible to find several configurations differing from the original one and having a higher efficiency with respect to the criterion of accuracy and speed of calculations. One of these configurations is proposed below for the use in the optimized parametrization of the term $Nl(\sigma, \theta)$.

Quantitatively, the advantages of the optimized version of FDIA are as follows:

(i) The calculation speed is increased twofold for the term Nl alone [13]; as a result, the time needed for WAM calculation of the term Nl takes only about 30% of the entire time of wave forecasting.

(ii) The relative error of the approximation of the exact calculation of Nl is about 40%, i.e., 1.5 times less than the error for the original version of the DIA [18].

In the new model, the use of the optimized version of the FDIA mentioned above is suggested.

The algorithm of estimating $Nl[S(\sigma, \theta)]$ by the optimized scheme lies in the following. The numerical frequency–angle grid $\{\sigma, \theta\}$ is given by the standard formulas

$$\sigma(i) = \sigma_0 e^{i-1} \quad (0 \leq i \leq I), \qquad (9a)$$

$$\theta(j) = -\pi + j\Delta\theta \quad (0 \leq j \leq J). \qquad (9b)$$

Here, $\sigma_0$ is the lower edge of the frequency band, $e$ is the frequency exponential increment, $I$ is the number of frequencies considered, $\Delta\theta$ is the angular resolution in radians, and $J$ is the number of angles under consideration. For the resolution parameters, the following values are recommended:

$$e \leq 1.1, \quad \Delta\theta \leq \pi/12. \quad (10)$$

The values of $\sigma_0$ and $I$ are chosen according to the conditions of solution of the problem. Typical values of other parameters can be $\sigma_0 = 2\pi 0.05$ rad/s, $I = 25$–$30$.

In the case when

$$e = 1.05 \text{ and } \Delta\theta = \pi/18, \quad (11)$$

an optimal configuration of four interacting waves in the version of the FDIA is given by the following relations:[2]

(a) For frequencies,

$$\sigma_1 = \sigma e^4, \quad \sigma_2 = \sigma e^5, \quad \sigma_3 = \sigma e^8. \quad (12a)$$

(b) For angles,

$$\theta_1 = \theta + 2\Delta\theta, \quad \theta_2 = \theta + 2\Delta\theta, \quad \theta_3 = \theta + 3\Delta\theta. \quad (12b)$$

Here, $\sigma$ and $\theta$ are considered to be known. (As was mentioned above, these are rotated within the numerical loop.) In the remaining part, the process of estimating the term $Nl[S(\sigma, \theta)]$ is determined by typical relations for the DIA of form (7) and (8). The fitting constant $C$ is 12 000 for the optimized version of the FDIA in the full version of the new source function.

Thus, the algorithm of calculating an optimized version of the nonlinear term of the source function in the frequency–angle representation for the wave energy spectrum $Nl[S(\sigma, \theta)]$ is completely determined.

## 3. PARAMETRIZATION OF THE ENERGY-INPUT MECHANISM (IN TERM)

### 3.1. General Formulation

Theoretical grounds for representation of the input term in a spectral form were given in the pioneering papers by Phillips [19] and Miles [20] almost a half-century ago. Since that time, many authors have contributed to theoretical solution of the problem, but the final form of parametrization has not been found yet. Therefore, for input-term description, a semiempirical representation is commonly used [1, 2]

$$\text{In} = \beta(\sigma, \theta, U)\sigma S(\sigma, \theta), \quad (13)$$

which corresponds to Miles' approach.

The problem of describing the energy-input mechanism lies in specifying the form of wave-growth increment $\beta(\sigma, \theta, U)$ as a function of the parameters of the wave spectrum and local wind at the standard horizon of 10 m $\mathbf{U}(x, t) \equiv \mathbf{U}_{10}$. General theoretical models become inefficient owing to the extreme complexity of the problem of describing turbulent processes taking place on the wavy air–sea interface. Therefore, for better grounding the parametrization of the coefficient $\beta$, one needs to combine empirical data with some theoretical results derived by certain mathematical models for the boundary layer of the atmosphere [14, 21, 22]. One should keep in mind that theoretical models are sometimes more informative than direct experimental data because of the difficulty in gaining reliable experimental data in a wide band of wave frequencies.

Here, we will not consider numerous points of the problem mentioned but note only three principal aspects of coefficient $\beta$ parametrization. They are as follows:

the size of the frequency interval where the parametrization of input term is valid;

the existence of a frequency domain where the value of $\beta$ is negative;

the form of wind representation; that is, the wind at a certain fixed horizon (for example, $\mathbf{U}_{10}$) or the wind friction velocity $u_*$, given by the relation

$$u_* = C_d^{1/2}(z)U(z), \quad (14)$$

where $C_d(z)$ is the drag coefficient for the horizon $z$.

Only in the aspects mentioned will we compare different parametrizations, their merits, and efficiency with the criterion for the accuracy and speed of calculation. All other points, namely, the dependence of $u_*$ (or $C_d$) on wave age $A$, defined by the relation

$$A = c_p/U_{10} = g/\sigma_p U_{10} = \tilde{\sigma}_p^{-1} \quad (15)$$

($\sigma_p$ is the peak frequency of a wave spectrum), and other questions related to the point (for example, dependence of $C_d$ on wind speed $U(z)$), will not be discussed here because of the high degree of their uncertainty (see below). This is our fundamental modification to the conventional approach.

To confirm the above thesis about uncertainty, it is sufficient to demonstrate the plots from [23, 24], which testify to a fundamental uncertainty for the dependence of the drag coefficient on the wind speed $C_d(U)$ (Fig. 2) and the normalized roughness height (Charnock's parameter) on the inverse wave age $\tilde{z}_0 = z_0 g/u_*^2$ (Fig. 3). The interpretation of the uncertainty mentioned above has been discussed in [11] recently, which permits us to leave this point aside, referring the readers to the cited literature.

### 3.2. Comparative Analysis

In the context of the three above aspects of the parametrization of the coefficient $\beta$, let us make a comparative analysis of the widely used approximations.

---

[2] The constants in relations (12a) and (12b) for the optimized version of the FDIA depend on $e$ and $\Delta\theta$. This is easily controlled by instructions for the method.

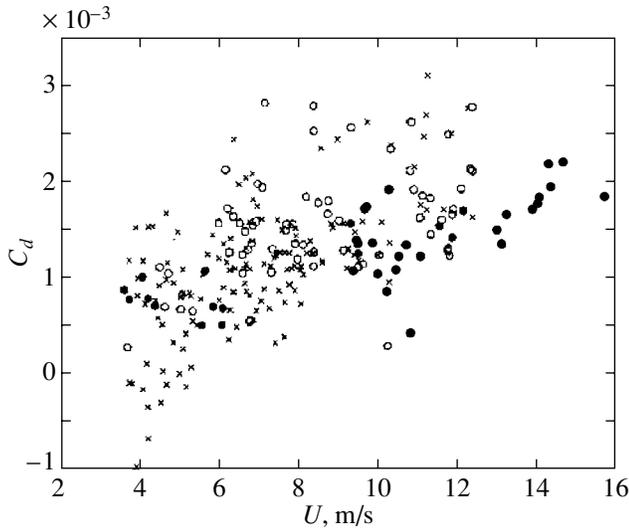

**Fig. 2.** Variation of the drag coefficient $C_d$ vs. the wind $U_{15}$ for different wave-origin conditions according to [23].

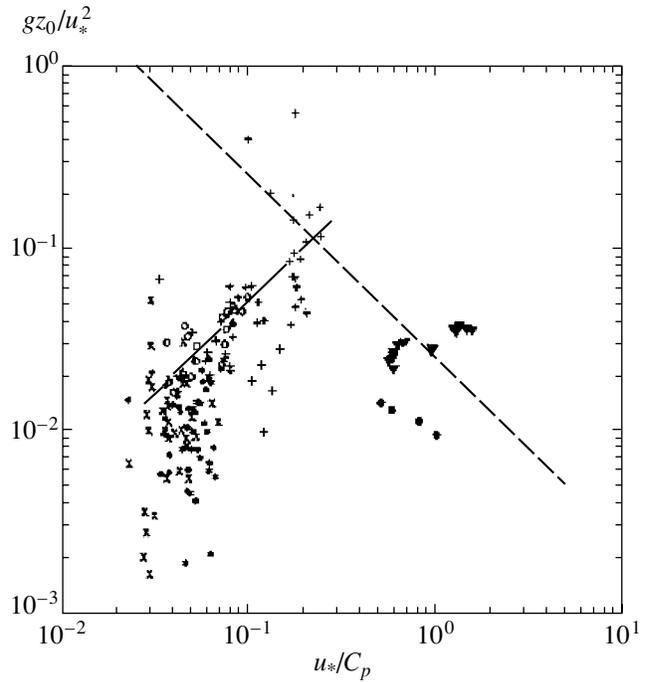

**Fig. 3.** Variation of the Charnock's parameter vs. the inverse wave age, following [24]. The straight lines are some of the simplest empirical parametrizations.

For example, in the WAM model [4], they use an empirical parametrization proposed by Snyder *et al.* [25] more than twenty years ago, which has the form

$$\beta(\sigma, \theta, \mathbf{U}) = \max\left[0, a\left(\frac{28 u_* \sigma}{g}\cos(\theta - \theta_u) - b\right)\right]. \quad (16)$$

Here, $\theta_u$ is the local wind direction and the friction velocity $u_*$ is recalculated permanently, in the aspect of dependence on the wave age $A$, by some semiempirical manner according to Janssen's theory [22]. (The values of the fitting parameters $a$ and $b$ are not important in our consideration.) The main shortage of parametrization (16) is an extremely small frequency domain of its validity. The authors of [25] themselves postulated the domain of validity for parametrization (16) by the relation

$$1 \leq U_5 \sigma / g \leq 3, \quad (17)$$

which evidently does not satisfy the requirements for problems of wave calculation. For this reason, the approach used in the widespread version of WAM needs a definite improvement.

Such an improvement of empirical parametrization of $\beta$ ($\sigma$, $\theta$, $U$), combining the experimental results of [25, 26], was carried out in 1987 [27]. It has the form

$$\beta = \max\left\{0, \left[0.04\left(\frac{u_* \sigma}{g}\right)^2 \right.\right.$$
$$\left.\left. + 0.00544\frac{u_* \sigma}{g} + 0.000055\right]\cos(\theta - \theta_u) - 0.00031\right\} \quad (18)$$

and is valid in the frequency range

$$1 \leq U_5 \sigma / g \leq 75, \quad (19)$$

which is acceptable for solution of all practical problems of wind-wave modeling. The use of (18) in some particular versions of WAM [28] counts in favor of such a change. The fundamental advantage of parametrization (18) for the coefficient $\beta$ lies in its closeness to the theoretical result by Chalikov [21], who found from numerical experiments that a theoretically substantiated formula for $\beta$ should have the form

$$\beta = 10^{-4}(a\tilde{\sigma}^2 + b\tilde{\sigma} + c), \quad (20)$$

where the fitting parameters $a$, $b$, $c$ should be calculated in a rather complicated manner via estimation of the normalized frequency variable $\tilde{\sigma}$ (for details, see [6]). Such a parametrization of $\beta$ is used in the WW model [6], which is a more recent model than WAM.

The second important point of the theoretical results by Chalikov lies in the proof of the existence of negative values of $\beta$ in the frequency domain $U\sigma/g \leq 1$, where waves are ahead of the local wind. In the WAM model, this physically evident fact is ignored absolutely, but it is accepted in the WW model. Our experience in the process of fitting the wind-wave model shows that taking this physical effect into account does play a significant role. Consequently, it must be accepted in any present-day model.

It follows from formal logic that it is reasonable now to use the parametrization of $\beta$ accepted in the WW

model. However, today, it is well known that excessive details in the calculation of parameters in (20) lead to a significant slowdown of wave forecasting by the WW model (twofold compared to WAM) (for example, see [29]). Therefore, parametrization (20) should be optimized.

We propose the following parametrization of $\beta$ of form (20), optimized using empirical relation (18) and theoretical results by Chalikov:

$$\beta = \max\left\{-b_L, \left[0.04\left(\frac{u_*\sigma}{g}\right)^2 + 0.00544\frac{u_*\sigma}{g} + 0.000055\right]\cos(\theta - \theta_u) - 0.00031\right\}. \quad (21)$$

As is seen, in addition to (18), parametrization (21) includes the condition of existence of negative values of $\beta$ with the limiting magnitude

$$b_L = 0.000005. \quad (22)$$

The order of $b_L$ follows from theoretical estimations [21], but the exact magnitude is found by fitting the model against empirical wave-growth curves (see Section 5).

It remains to determine how to use the transition from the wind, $U_{10}$, to the friction velocity $u_*$.

### 3.3. Choice of Wind Representation

In the case of a constant drag coefficient $C_d$, there is no fundamental difference between the representations of $\beta$ in terms of $U_{10}$ and of $u_*$. It is more complicated to take into account the dependence of $C_d$ on wind and parameters of wave state: the wave age $A$, the shape of spectrum $S(\sigma, \theta)$, etc. The question arises as to what extent these complicated dependences can be reliably taken into account in practical calculations.

In this aspect, in such widely used models as WAM and WW, simplified forms of the dependence of $C_d$ on the system's parameters indicated above are used. However, as we already noted, all existing simplified dependences (with one or two parameters) cannot be substantiated theoretically because all of these are rough empirical simplifications. This point was studied in detail in our special paper [11], where it was shown that correct interpretation for variability of the characteristics of air–sea interface can be given only on the basis of a physical model for the boundary layer similar to that proposed in [14].

Indeed, analysis of Fig. 2 shows that there is generally no unambiguous dependence $C_d(U_{15})$. Moreover, as one can see from Fig. 3, the scatter in the values of the normalized height of roughness $\tilde{z}_0 = z_0 g/u_*^2$ can reach several orders of magnitude at any fixed wave age a phenomenon, which fundamentally defies a simple interpretation. At the same time, the use of a physical model for the boundary layer proposed in [14] allows a reasonable interpretation simultaneously for all the data mentioned above [11]. However, for this purpose, it is necessary to know the shape of the two-dimensional spectrum of waves $S(\sigma, \theta)$ up to frequencies of about 10–15 Hz.

It follows that $U_{10}$ should be recalculated to $u_*$ either by the full model of the dynamic boundary layer [11] or by using simplifications of the assumption that $C_d$ is constant. Our calculations show that, on the time scale of wave development (several hours), this assumption is not worse than the simple parametrizations of $C_d(U_{10}, A)$ used in WAM and WW from the standpoint of wave forecasting.

Thus, with the criterion of accuracy and speed of calculation, the combination of the two following conditions is optimal: the constancy of the coefficient $C_d$ on a scale of several hours of wave evolution and a periodic correction of this coefficient with the model of [14].[3] In this case, a possibility appears for a regular recalculation of all boundary-layer characteristics as functions of the wave stage (including the wind profile $U(z)$ in the limits of the standard horizon 0 m $< z <$ 10 m). Such a wind-wave model acquires a new character that qualifies it as a model of the next, that is, fourth, generation, which is conventionally referred to as the wave model with the dynamic boundary layer. Thus, the problem of optimization for the input-term parametrization in the source function of the numerical wave model is entirely solved in the context of the problem formulated above.

### 3.4. Dynamic Boundary-Layer Block

For the completeness of presentation of the source function algorithm, let us rewrite the main formulas for the dynamic boundary layer, following [14].

According to [14], the wind profile averaged over a long time compared to the wave period is calculated by the formula

$$U(z) = \frac{u_*}{0.4}\int_{z_0^v}^{z}\left[1 - \frac{I(z)}{1 + I(0)}\right]^{3/4} d(\ln z) = \frac{u_*}{0.4}J(z), \quad (23)$$

where $z_0^v$ is the viscous sublayer height on the order of

$$z_0^v \cong 0.1\frac{\nu}{u_*^t} \cong 0.00005/U_{10}. \quad (24)$$

---

[3] As the initial value for the friction velocity, the relation $u_* = U_{10}/26$ is acceptable.

In (23), the quantity $I(z)$ is determined by the integral over the wave spectrum shape as

$$I(z) = \int_{\sigma_{min}}^{\sigma_{max}} d\sigma \oint_\theta d\theta \left[ \exp(-zk/\pi) \right.$$

$$\left. \times \cos(5\pi zk) \beta\left(\frac{u_*\sigma}{g}, \theta\right) k^2 S(\sigma, \theta) |\cos(\theta)| \right], \quad (25)$$

where $\beta\left(\frac{u_*\sigma}{g}, \theta\right)$ is given by formula (21) and $\sigma_{max}$ is the upper limit of integration, which has an order of 70–90 rad/s and is chosen during fitting of the model. The lower limit $\sigma_{min}$ is evidently defined by the lower boundary of the frequency band. In the frequency domain above the higher edge of the numerical frequency grid, the high-frequency spectral tail is parametrized on the basis of special observational data (see [14]). According to up-to-date information about uncertainty for the law of spectral-tail fall [30], we adopted the hypothesis that the frequency–angle spectrum has the form

$$S(\sigma, \theta) \sim \sigma^{-5} \cos^2(\theta - \theta_u). \quad (26)$$

As calculations show, the results of estimations for boundary-layer characteristics are weakly sensitive to variations in the angular shape very sensitive to variations in the frequency dependence. The elaboration of the parametrization of the spectral tail needs additional special investigation.

Finally, the boundary-layer parameters are determined by the formulas

$$u_* = 0.4 U_{10}/J_{10} \quad (27)$$

and

$$C_d(10) = [0.4/J_{10}]^2. \quad (28)$$

If the hypothesis of a logarithmic boundary-layer wind profile is accepted, the roughness height is determined by the relation

$$z_0 = 10/\exp(J_{10}). \quad (29)$$

The relations presented above give an insight into the character of operation of the wind-wave model in the regime of the dynamic boundary layer and close the description of the input term in the model proposed.

From direct calculations, it was established that, in the fitting regime of model operation, it is acceptable to use a constant drag coefficient given by the relation

$$u_* = U_{10}/26. \quad (30)$$

In a general case, the transition $u_* \Leftrightarrow U_{10}$ should be carried out using the block of the dynamic boundary layer of the atmosphere described above. This block could be either turned on or turned off in accordance with a specific formulation of the problem.

## 4. PARAMETRIZATION OF THE DISSIPATION MECHANISM (DIS TERM)

### 4.1. Initial Statements

The dissipation term is the least investigated both theoretically and experimentally. Despite remarkable efforts in this direction (for example, see [31–33]), a widely recognized mathematical description of the dissipation term in the source function has not been found yet.

In the WAM model, the following quasi-linear (in the wave spectrum) parametrization of Dis is used:

$$\text{Dis} = \gamma(\sigma, \theta, \mathbf{U}, E) \sigma S(\sigma, \theta), \quad (31)$$

whose spectral form was substantiated in [34] more than 30 years ago. It is suggested that, in a general case, the dimensionless fitting function $\gamma$ can depend on both wind $\mathbf{U}$ and wave energy $E$. However, in the WAM model, these dependences were found by the trial-and-error method, which has no physical content.

Criticism of parametrization (31) has continued for a long time (for example, see [35]). Therefore, we will not consider this point. Note only that the authors of [35] proposed a series of modifications of Dis based on the statistics of wave breaking, which showed their efficiency as applied in WAM [28, 36]. However, parametrizations proposed in [28, 35, 36] have a purely empirical nature and can hardly be substantiated theoretically.

A more thoughtful parametrization for the dissipation term is used in the WW model [6]. On the basis of the representation of dissipation due to viscosity, Tolman and Chalikov [6] have written a phenomenological expression for Dis in the form

$$\text{Dis} = -[u_* h(E) \phi(\alpha)] k^2 S(\sigma, \theta), \quad (32)$$

where $u_*$ is the friction velocity, $h(E)$ is the wave height as a function of wave energy $E$, and $\phi(\alpha)$ is the dimensionless fitting function depending on the spectrum-intensity parameter at high frequencies $\alpha$. Note that the expression in the square brackets has the dimension of the effective viscosity of the wave system under consideration. Owing to this fact, expression (32) is physically meaningful. However, further in [6], a system of hypothetical considerations is presented. The authors of [6] divided the frequency interval into three portions and assumed that the dissipation mechanism is different in different domains. In each domain, they fitted phenomenological expression for $\phi(\alpha)$. Next, they had to join the formulas obtained for each domain by highly arbitrary calculations. This arbitrariness in constructing the function $\phi(\alpha)$ in [6] for the parametrization of Dis in the WW model is the issue most vulnerable to criticism.

From the standpoint of theoretical substantiation, the concept of describing the mechanism of wave-energy dissipation through interactions of wave motions with the turbulence of the upper water layer is actually of the most interest. In such a case, the dissipation mechanism is equivalent to viscous losses; i.e., it admits a clear physical interpretation. In the spectral representation, the mechanism of viscous losses may be represented by the well-known theoretical formula

$$\text{DIS}(\sigma, \theta, S, \mathbf{U}) = \nu_T(\sigma, \theta, S, \mathbf{U}) k^2 S(\sigma, \theta), \quad (33)$$

where the function of effective viscosity $\nu_T$, in principle, can depend on the wave spectrum shape directly. Such a concept was proposed in [37] back in 1986. This concept was successfully used in a number of models [9, 38]. Later, the concept was substantiated theoretically in [39, 40]. The main statements of the above theory are as follows.

### 4.2. Theoretical Statements

The general statement on the dissipation mechanism is based on the concept of wave-motion energy losses due to the turbulence of the upper water layer. The nature and features of the upper-layer turbulence do not play any role. The nature of turbulence could be arbitrary: wave breaking, shear flows at the interface, sprinkling, foam, etc. The matter is that the field of potential wave motion, $\mathbf{V}_w(\mathbf{x}, t)$, interacts with the field of turbulent motion, $\mathbf{V}_t(\mathbf{x}, t)$. In such a case, after averaging the fluid-dynamic equations over the scales of turbulent motions, the Reynolds stresses inevitably arise, which cause the term of viscous losses to appear. If now the closure of the stresses is carried out through the wave-velocity variables $\mathbf{V}_w$, the resulting system of equations for waves can be treated by the conventional spectral method. In this manner, expression (33) can be derived (for details, see [39, 40]).

Further, it is very logical to suppose (and the closure hypotheses for Reynolds' stresses do permit this) that the desired function of effective viscosity $\nu_T$ is a series in powers of the wave spectrum $S(\sigma, \theta)$; i.e., it can be represented as

$$\nu_T = \sum_{n=0}^{N} c_n(\sigma, \theta, g, \mathbf{U}) S^n(\sigma, \theta). \quad (34)$$

With consideration for the presence of a small dimensionless parameter of the wind-wave system on the order of $\alpha = S(\sigma, \theta)\sigma^5/g^2 \cong 10^{-2}$, relation (34) can be rewritten as

$$\nu_T = \sum_{n=0}^{N} \nu_n(\sigma, \theta, g, \mathbf{U}) \alpha^n(\sigma, \theta), \quad (35)$$

where the coefficients $\nu_n(\ldots)$, having the dimension of viscosity, no longer depend on the wave spectrum. It is evident that, owing to the smallness of the parameter $\alpha$, it is sufficient to restrict ourselves to a few first terms in (35).

The zero-order term $\nu_0$ is not necessarily equal to zero. However, for a numerical wave model, this term is not important because it leads to the term of Dis that is linear in the spectrum. Such a term is "absorbed" by the linear term In (by fitting the factors in expressions (13) or (21)). Therefore, for numerical-model construction, the most important in (35) is the term linear in the wave spectrum, which corresponds to the simplest closure of Reynolds' stresses (see [39, 40]). Consequently, the desired expression for the dissipation term of the source function assumes the form

$$\begin{aligned}
\text{Dis}(\sigma, \theta, S, \mathbf{U}) &= \left[\nu_1(\sigma, \theta, g, \mathbf{U})\left(\frac{S\sigma^5}{g^2}\right)\right] k^2 S \\
&= \tilde{\gamma}(\sigma, \theta, \mathbf{U}) \frac{\sigma^6}{g^2} S^2(\sigma, \theta),
\end{aligned} \quad (36)$$

where the only undefined term is the dimensionless dissipation coefficient $\tilde{\gamma}(\sigma, \theta, \mathbf{U})$.

A general form of $\tilde{\gamma}(\sigma, \theta, \mathbf{U})$ is found using the existence condition for an equilibrium wave spectrum $S_{eq}(\sigma, \theta)$ in the high-frequency region conventionally defined by the relation

$$\sigma > 2.5\sigma_p. \quad (37)$$

The existence of equilibrium means that, for the aforementioned spectral shape, the balance of terms in the source function is close to zero; i.e.,

$$F|_{S=S_{eq}} = [\text{Nl} + \text{In} - \text{Dis}]|_{S=S_{eq}} \approx 0. \quad (38)$$

If one takes into account that the relative contribution of Nl in the high-frequency region is on the order of 10%, relation (38) can be rewritten as

$$[\text{In} - \text{Dis}]|_{S=S_{eq}} \approx 0. \quad (39)$$

If the parametrization of the term In is known, the desired function $\tilde{\gamma}(\sigma, \theta, \mathbf{U})$ and the dissipation term as a whole valid in the spectral-tail domain (37) can be obtained unambiguously from (39). An analytical extension of the formula obtained for Dis to the energy-containing domain of the spectral peak is carried out by a phenomenological fitting in the course of testing calculations. Thus, the semiphenomenological theoretical substantiation of the parametrization of the Dis term in the form of (36) is completed.

To conclude this subsection, it remains to note that the choice of a specific dependence $\tilde{\gamma}(\sigma, \theta, \mathbf{U})$ leaves a certain arbitrariness that admits the choice of the equilibrium spectrum $S_{eq}(\sigma, \theta, \mathbf{U})$ as a function of its parameters. This arbitrariness is justified by some degree of uncertainty in the form of the spectrum $S_{eq}(\sigma, \theta, \mathbf{U})$, as follows from analysis of experimental data [30]. Nevertheless, a general form of parametrization (36), being theoretically substantiated, remains robust to any uncertainties in the shape of $S_{eq}(\sigma, \theta, \mathbf{U})$. This permits us to hope for a universal character of this term for applications in numerical models for wind waves.

### 4.3. Specification of the Dis Term

Taking into account that the shape of the equilibrium frequency spectrum is given by (26), we propose the use of the following expression for the dissipation term:

$$\text{Dis}(\sigma, \theta, S, \mathbf{U}) = c(\sigma, \theta, \sigma_p) \\ \times \max[0.00005, \beta(\sigma, \theta, \mathbf{U})] \frac{\sigma^6}{g^2} S^2(\sigma, \theta). \quad (40)$$

Here, the value of $\beta(\sigma, \theta, \mathbf{U})$ is given by formula (21) derived earlier and a dimensionless fitting function $c(\sigma, \theta, \mathbf{U})$, describing fine details of dissipation processes in the vicinity of the peak frequency $\sigma_p$, is

$$c(\sigma, \theta, \sigma_p) = 32\max[0, (\sigma - 0.5\sigma_p)/\sigma] T(\sigma, \theta, \sigma_p), \quad (41)$$

where the angle-dependent function has the form

$$T(\sigma, \theta, \sigma_p) \\ = \left\{ 1 + 4\frac{\sigma}{\sigma_p}\sin^2\left(\frac{\theta - \theta_u}{2}\right) \right\} \max[1, 1 - \cos(\theta - \theta_u)]. \quad (42)$$

Here, the standard notation $\max[\ldots, \ldots]$ means the choice of the maximum quantity among the quantities between the brackets.

As follows from (40), in the proposed parametrization for Dis, in addition to other features, the condition is imposed that a nonzero minimum level of energy losses exists for waves at frequencies on the order of or less than $\sigma_p$. This condition reflects background viscous dissipation processes occurring in reality. Note that this feature is introduced for the first time in the parametrization of the Dis term.

## 5. PRINCIPLES AND METHODOLOGY FOR TESTING NUMERICAL MODELS

### 5.1. Principles of Testing and the List of the Main Tests

For the first time, regular testing problems for numerical wind-wave models were proposed in implementing the SWAMP project [3]. The authors of the project restricted themselves to a comparison of the results of testing a number of models. In our view, the testing process has a more fundamental meaning because it is an integral element for the technology of construction of any numerical model, an aspect which is often overlooked (for example, see [2, 8]). In this respect, it is pertinent to give the necessary motivation and specification for the process of testing, the basis of which was formulated in [1]. A short presentation of testing principles is given below.

First, the main aim of testing is to obtain fundamentally important quantitative and qualitative information about general features of a given wind-wave model. The presence of well-controlled conditions for wave formation provides an essential advantage of such academic tests over a model verification from full-scale data, when the wind field is available with a large degree of uncertainty.

Second, testing problems should be informative and predictable. A measure of a test's "informative value" is the scatter in the so-called control (or "informative") parameters about empirically observed or theoretically expected values estimated on the basis of general physical considerations. The greater the scatter in these values obtained for different models, the more informative the test. The choice of control parameters is an integral part of the formulation of testing problems.

Third, the testing problem should be sufficiently simple and sharply aimed. This requirement makes it possible (a) to assess the role and properties of each part of the model maximally independently of one another, (b) to reveal integral properties of a model as a whole and distinguish models by the quantitative characteristics of description of major elements of wave evolution (these elements include processes, such as wave growth, swell decay, the turning of wave into the wind, interactions of the elements of the wave system, etc.); and (c) to determine the degree of reproducibility of a model of the observed effects and reveal new physical effects in wind-field evolution. In this respect, simple academic testing problems are very important from the physical point of view.

These testing principles could be generalized to more complicated cases: nonstationary wind fronts, hurricanes, etc.; however, this issue is beyond the scope of this study. For this reason, we formulate and consider several testing problems alone that have shown their high informative merits [1–3]. According to the conclusions made in [1], these are as follows:

Test 1. Straight fetch (wave development or tuning test).[4]

Test 2. Swell decay (dissipation test).

Test 3. Swell along the wind (swell–wind wave interaction test).

Test 4. Turning of the wind.

Test 5. Diagonal wind front.

Note that tests 1, 4, and 5 are known as the standard SWAMP tests [3]. Tests 2 and 3 were proposed by me together with Efimov in constructing the models of the second and third generation [1]. For the dynamic boundary block, some problems considered in [11] can play the role of tests. However, here, we will not consider them, referring the readers to the cited literature.

---

[4] Usually, this test is used to tune the fitting parameters in the source function of the model.

In addition to the aforementioned, one should note the following. In view of an announcing character of this paper and for reasons of space, it is impossible here to consider the results of complicated tests, such as tests 4 and 5, because of a certain degree of uncertainty of their treatment [1, 3, 6, 9]. For this reason, their discussion is beyond the scope of this paper.

*5.2. Control Quantities and Levels of Model Adequacy*

Before formulating the testing problems, it is necessary to define the desirable degree of detail in describing the phenomena of interest for any numerical model. Here, we touch on the methodical point of model classification by the level of their adequacy, which was considered earlier for a general case in [1]. Owing to its importance, it is pertinent to consider this point in the aspect of its application to wind-wave modeling because this point is closely connected to the choice of control quantities for testing tasks.

It is evident that, the more details of the field under study that are reproduced by the model, the higher the level of model adequacy. For each level of detail, it is suggested that reference empirical data be available and the degree of their accuracy be known. As for wind waves, the most informative representation of the wave field is the spatial distribution of the two-dimensional wave spectrum $S(\sigma, \theta, \mathbf{x}, t)$ (the highest level of adequacy). By a simple integration, information for the one-dimensional spectrum $S(\sigma, \mathbf{x}, t)$ and for the parameters of its shape can be found (middle level of adequacy). Then, the integral values of the wave field such as the wave energy $E(\mathbf{x}, t)$, the peak frequency $\sigma_p(\mathbf{x}, t)$, and the general wave-propagation direction $\theta_p(\mathbf{x}, t)$ can be found (the lowest, first level of adequacy). All the other physical characteristics of the wave field, which can be calculated from those listed above (the mean wave height, main period, etc.), correspond to the lowest level of adequacy. For simplicity, below, we will restrict ourselves to analysis of stationary wave fields, when the dependence on time disappears.

The aforementioned classification can be applied only to test 1, for which there are generally recognized empirical dependences of integral values of the wave field on the fetch X (characteristics of the first level of adequacy). For this test, there is also information on some characteristics of the one-dimensional spectrum and on a few characteristics of the two-dimensional spectrum (characteristics of the next levels of adequacy). Therefore, it should be remembered that all control quantities can be substantiated only for test 1. For the other tests, the choice of control quantities is possible at an expert level only, which means the choice at the level of one's experience of numerical simulation and physical intuition.

Thus, for test 1, the following control quantities correspond to the first level of adequacy: the wave energy $E(\mathbf{x})$ and the peak frequency $\theta_p(\mathbf{x})$ (the value of $\sigma_p(\mathbf{x})$ is excluded owing to the trivial feature of the problem). For more generality, these control quantities of the test are considered in dimensionless units: the dimensionless energy $\tilde{E} = \dfrac{Eg^2}{U_{10}^2}$ and the dimensionless peak frequency $\tilde{\sigma}_p = \dfrac{\sigma_p U_{10}}{g}$ are treated as functions of the dimensionless fetch $\tilde{X} = \dfrac{Xg}{U_{10}^2}$. For the reference empirical dependences, two types of dependences can be recommended at the present time [2]:

(a) for a stable atmospheric stratification,

$$\tilde{E}(\tilde{X}) = 9.3 \times 10^{-7} \tilde{X}^{0.77}, \quad (43)$$

$$\tilde{\sigma}_p(\tilde{X}) = 12 \tilde{X}^{-0.24} \text{ and} \quad (44)$$

(b) for an unstable atmospheric stratification,

$$\tilde{E}(\tilde{X}) = 5.4 \times 10^{-7} \tilde{X}^{0.94}, \quad (45)$$

$$\tilde{\sigma}_p(\tilde{X}) = 14 \tilde{X}^{-0.28}. \quad (46)$$

We note that these empirical dependences are valid only in the interval of fetches $100 < \tilde{X} < 10000$, and statistical errors of parameters in them are about 10–15% [2].

To check the second level of adequacy, the following control characteristics can be used:

(a) the power $n$ obtained for the law of spectral-tail fall at high frequencies

$$S(\sigma) \propto \sigma^{-n}, \quad (47)$$

(b) the level of tail intensity for the one-dimensional spectrum defined by the relation

$$\alpha(\sigma) = S(\sigma)\sigma^5 g^{-2}, \quad (48)$$

(c) the frequency width of the spectrum defined by the formula

$$B = \int S(\sigma) d\sigma / S(\sigma_p) \sigma_p, \quad (49)$$

and a series of other less important parameters of the one-dimensional spectrum shape.

For a control quantity for the third level of adequacy, one may use the angular-narrowness function for the two-dimensional spectrum given by the formula

$$D(\sigma) = S(\sigma, \theta_p) / \int S(\sigma, \theta) d\theta. \quad (50)$$

For completeness, one should note that, as the referring dependences, it is possible to use some observable evolution effects as well (for example, the "overshoot" effect). However, because of the lack of space, this issue is omitted here.

All the quantities enumerated above have the known empirical values (at least for a fully developed wind sea), thus permitting quantitative estimations of the degree of correspondence of any model to full-scale data. However, we will not consider this issue in this paper. At the stage of announcing the new model, we restrict ourselves mainly to the characteristics of the first level of adequacy and to a cursory mention of characteristics of the second level, while leaving more detailed comparisons for further studies.

## 6. FORMULATION OF TESTING PROBLEMS, RESULTS OF SIMULATIONS, AND ANALYSIS

### 6.1. Method of Numerical Simulations

Wind wave simulations were based on the numerical solution of Eq. (1) on the frequency–angle grid described by (11) at a number of frequencies $I = 36$ and a number of angles $J = 36$. The spatial grid included $NX$ nodes along the $OX$ axis, which is directed along the wind (a typical value is $NX = 30$), and $NY$ nodes in the transverse direction (for specification, see below). The source function was represented by the formulas given in Sections 2–4. The numerical solution of Eq. (1) was carried out by an implicit numerical scheme of directed differences along a flux having the first order of accuracy [1].

The initial wave field was specified by a spatially homogeneous two-mode wave spectrum in the following representation:[5]

$$S(f, \theta) = R1 S_1(f, \theta, f_{p1}, \theta_{p1}, \gamma_1, s_1) \\ + R2 S_2(f, \theta, f_{p2}, \theta_{p2}, \gamma_2, s_2), \quad (51)$$

where the coefficients $R1$ and $R2$ are responsible for the ratio of the intensities of the modes. The wind-sea mode $S_1$ was described by the typical JONSWAP spectrum [1, 2]

$$S_1(f, \theta, f_p, \theta_p, \gamma, s) = \alpha g^2 2\pi \sigma^{-5} \\ \times \exp(-1.25(f_p/f)^4) \gamma_J^{\exp[-(f-f_p)^2/0.01 f_p^2]} \Psi(s, \theta, \theta_p). \quad (52)$$

The swell mode $S_2$ was specified for each problem formulation in its own way (see below). In (52), $\alpha$ is the Phillips coefficient, $f_p = \sigma_p/2\pi$ is the peak cyclic frequency, $\gamma_J$ is the peak enhancing parameter ranging from 1.0 (for the Pierson–Moskowitz spectrum) to 3.3 (for the JONSWAP spectrum). The angular spread function had the form

$$\Psi(s, \theta, \theta_p) = I_s \cos^s(\theta - \theta_p), \quad (53)$$

---

[5] In simulations, it is preferable to use the cyclic frequency $f = \sigma/2\pi$ instead of the angular frequency $\sigma$.

with the normalization coefficient $I_s$ taken equal to 1 for simplicity of spectral shape variation. The general wave direction $\theta_p$ was, as a rule, directed along the local-wind direction $\theta_u$. From [1, 2], it is known that the initial spectral shape is, as a rule, not important in numerical simulations because it is replaced rapidly by current numerical wave spectra (exclusions are swell-decay simulations). Therefore, the specification of parameters for formula (52) is not discussed here.

For tests 1, 2, and 3, the boundary conditions at $X = 0$ were always invariant in time. At the final boundary $X = NX^*\Delta X$, the conditions were changed in accordance with calculations. The conditions at the lateral boundary were taken to be changeable in accordance with the change of the spectrum at internal grid points (so-called "fluid boundary conditions" or "open boundaries"). In such cases, to provide imitation for an infinite shoreline along the $OY$ axis, it is sufficient to use $NY = 3$, i.e., only three nodes along this axis [1]. The spatial steps were taken equal to each other, $\Delta X = \Delta Y$, but their values varied widely, depending on the formulation of the problem.

For tests 4 and 5, the boundary conditions at all the boundaries change in accordance with calculations but the values of $NX$, $NY$, and $\Delta X = \Delta Y$ correspond to the typical spatial sites used in [3].

### 6.2. Straight Fetch Test

**Formulation of the problem.** A spatially homogeneous and constant (in time) wind $\mathbf{U}_{10} = \text{const}$ blows normally to an infinite straight shore line ($X = 0$). The initial conditions are given by a spectrum of form (51), (52) with the following values of its parameters: $R1 = 1$, $R2 = 0$, $\alpha = 0.01$, $f_p \geq g/\pi U_{10}$, $\theta_p = 0$, and arbitrary values of $\gamma_J$ and $s$. The boundary conditions correspond to those mentioned in Subsection 6.1.

The main purpose of the test is to check the correspondence of stationary numerical dependences $\tilde{X}(\tilde{E})$ and $\tilde{f}_p(\tilde{X})$ to empirical growing curves (43), (44) and (45), (46) for different wind values. Additionally, one may estimate the adequacy level of the model by checking the degree of detail in describing the spectrum shape and evolution effects.

The results of model testing for wind values $U_{10} = 10, 20, 30$ m/s are presented in Figs. 4–7. From Figs. 4 and 5, it can be seen that numerical dependences $\tilde{X}(\tilde{E})$ and $\tilde{f}_p(\tilde{X})$ are in good agreement with the empirical wave-growth curves in the stable-stratification case within the limits of observation accuracy. Marked deviations from the empirical curves at some points close to the initial boundary (and splitting of the curves in the domain near $\tilde{X} \approx 10^3$ provided by the change of the discrete $\Delta X$) are an inevitable consequence of defects of the numerical method used to solve Eq. (1) (see the remarks in [6]). The limiting values of wave energy for a fully developed sea $\tilde{E}_d$ realized in the domain of very large fetches $\tilde{X} > 10^4$ correspond to the known empirical magnitude for the Pierson–Moskowitz spectrum: $\tilde{E}_d \approx (2-3) \times 10^3$ [1]. Therefore, one may state that the coefficients in the source function are fitted correctly, and the model has at least the first level of adequacy.

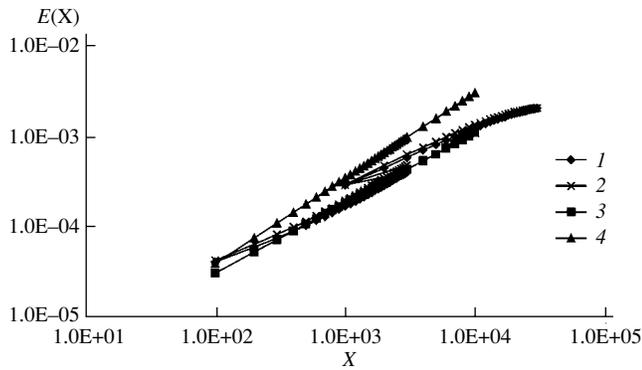

**Fig. 4.** Dependence of the dimensionless energy $E$ on the dimensionless fetch $X$: (*1*) wind $U_{10} = 10$ m/s, (*2*) wind $U_{10} = 30$ m/s, (*3*) empirical dependence (43), and (*4*) empirical dependence (45).

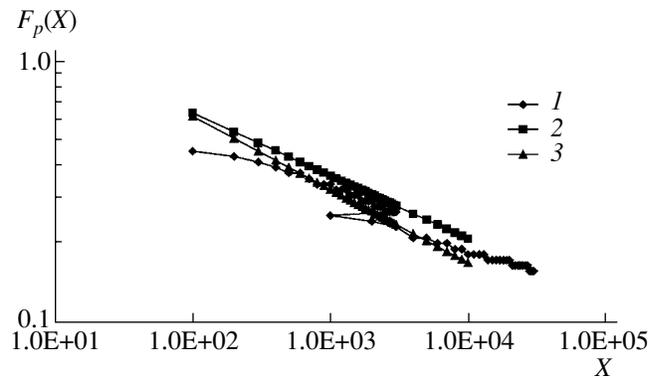

**Fig. 5.** Dependence of the dimensionless frequency $F_p$ on the dimensionless fetch $X$: (*1*) wind $U_{10} = 10$ m/s, (*2*) empirical dependence (44), and (*3*) empirical dependence (46).

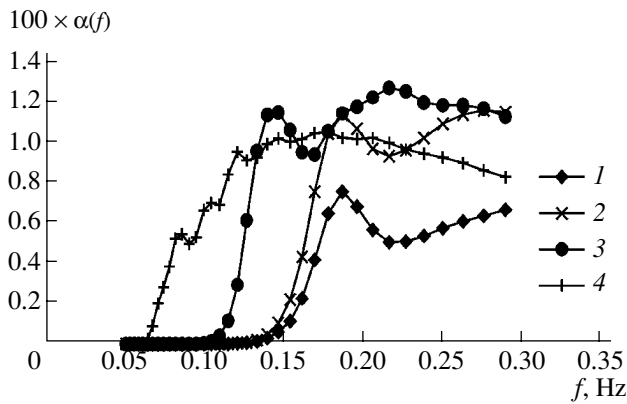

**Fig. 6.** Dependence of the intensity parameter on frequency, $\alpha(f) = S(f)f^5 g^{-2}$, at the boundary point of the calculation site for the four dimensionless moments $T = tg/U_{10}$: (*1*) $T = 0$, (*2*) $T = 10^3$, (*3*) $T = 10^4$, and (*4*) $T = 10^5$ ($U_{10} = 20$ m/s).

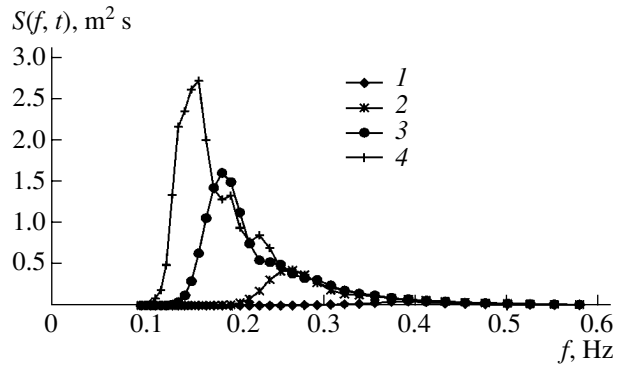

**Fig. 7.** Time history for the one-dimensional spectrum $S(f, t)$ at the boundary point of calculation site for four dimensionless moments $T = tg/U_{10}$: (*1*) $T = 0$, (*2*) $T = 10^4$, (*3*) $T = 3.3 \times 10^4$, and (*4*) $T = 10^5$ ($U_{10} = 10$ m/s).

Correspondence to the second level of adequacy is checked by a comparison of numerical characteristics for the one-dimensional spectrum (47)–(49) with their empirical values. Calculations show that, in the course of establishment of the developed rough sea, the following values of control quantities are realized in the model: $n \approx 4.5–4.9$, $B \approx 0.36–0.42$, and $\alpha \approx 0.008–0.01$. All of these magnitudes correspond well to those known from observations [1, 3, 8, 9, 30]. In addition, the model reproduces the effect of reducing the intensity of the spectral tail in the course of wave development (Fig. 6) and the overshoot effect (Fig. 7). It fol-

lows that the level of adequacy of the model discussed is no lower than the second level.

A more detailed description of model features in the aspect of estimating its degree of adequacy needs a separate presentation.

### 6.3. Swell-Decay Test

**Formulation of the problem.** The forcing wind is absent: $U_{10} = 0$. The initial wave state is given by a homogeneous wave field with the spectrum of form (51), (52) with parameters: $R1 = 1$, $R2 = 0$, $\alpha = 0.01$, $\gamma_J = 1$, $s = 2$, and $\theta_p = 0$.[6] The initial peak frequency $f_0 \equiv f_p(0)$ is fixed in the interval 0.08–0.32 Hz (which corresponds roughly to the spectra of a fully developed sea for winds from 20 to 5 m/s). Two or three runs are considered for different values of the peak frequency $f_0$. The boundary conditions correspond to those in Subsection 6.1.

---

[6] This spectrum of the PM type is not typical of the ordinary swell, which has a smaller intensity and a spectral tail of the form $S(f) \sim f^{-n}$ with $n \approx 6–7$ (normal swell) [1]. Therefore, in this text, the swell can be conventionally called a "strong swell," while retaining the term "weak swell" for the case of a special model spectrum with $n = 10$ (see [1]).

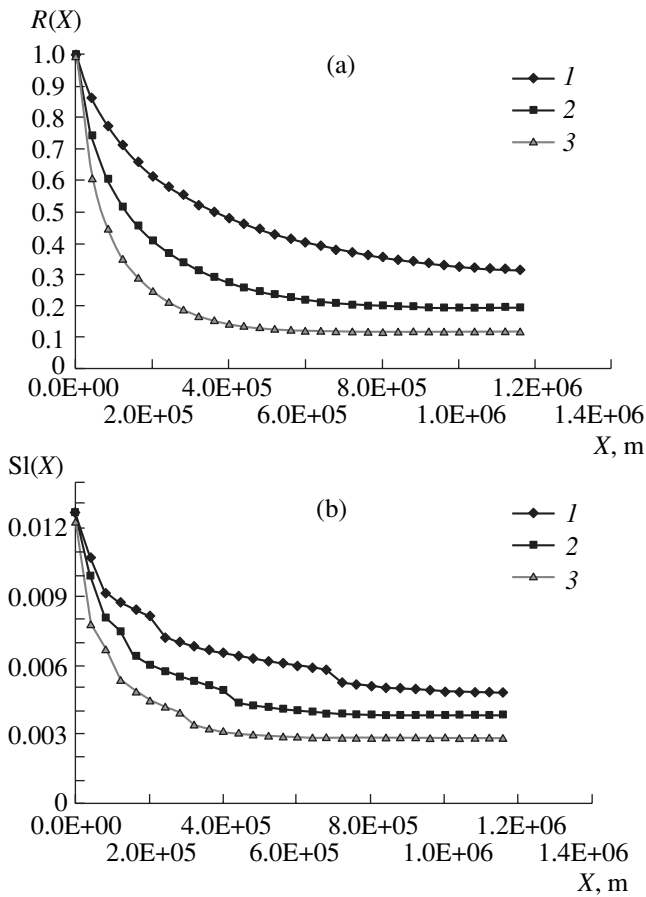

**Fig. 8.** Dependence of swell characteristics on the propagation distance according to test 2 for three initial peak frequencies: (*1*) $f_0 = 0.08$ Hz, (*2*) $f_0 = 0.15$ Hz, and (*3*) $f_0 = 0.24$ Hz. (a) The relative swell-energy reduction function $R(X)$ and (b) the mean wave slope $Sl(X)$.

The main aim of the test is to reveal quantitative features inherent in the model for the process of swell decay with distance for different initial peak frequencies $f_0$. Additionally, it is possible to investigate quantitative laws for spectral-shape changes during swell decay.

It should be noted that reliable empirical data on the laws of swell decay are virtually unavailable [1, 2, 7, 8]. Therefore, the control quantities are chosen at an expert level (for explanation, see above in Section 5). The following quantities are proposed as the control values:

(a) the relative energy losses in relation to the distance of propagation $X$,

$$R(X) = E(X)/E(0), \text{ and} \quad (54)$$

(b) the wave slope in relation to the distance $X$,

$$Sl(X) = H_m(X)/\lambda_p(X)$$
$$= [2E(X)]^{1/2} 2\pi (f_p(X))^2/g. \quad (55)$$

Here, $H_m$ is the mean wave height, $\lambda_p$ is the wavelength corresponding to the peak frequency, $X$ is the distance of swell propagation from the boundary $X = 0$.

For the characteristics of the spectral shape, the parameters given by relations (47)–(50) are of interest, both in their relation to the distance and as determined at a certain control point. However, because of the lack of space, these characteristics are not considered in this paper.

Results of testing are shown in Figs. 8a and 8b. Analysis of this figure leads to the following conclusions:

(i) The rate of swell-energy dissipation depends strongly on the initial peak frequency $f_0$: the greater $f_0$, the stronger the swell decay.

(ii) At a small distance of propagation ($X \leq 10^3 \lambda_0$), the character of "strong-swell" decay has an exponential feature; i.e., the dissipation of swell is accompanied by sharply diminishing the rate of decay per propagation-distance unit. At large distances ($X > 10^4 \lambda_0$), the rate of swell decay becomes close to zero (Fig. 8a).

(iii) During swell propagation, the peak frequency is shifted to the long-wavelength domain owing to a nonlinear mechanism of evolution. For this reason, the change of the mean slope $Sl(X)$ has a jump like character (owing to a discrete feature of the frequency set) (Fig. 8b).

(iv) Since the wave slope is a complex parameter of the system, this quantity is preferable for obtaining a qualitative description of dissipation features of the model. In [1], a criterion was proposed for such a description, which is based on the fixing of the length of swell propagation $L(f_0)$ at which the slope is diminished twofold. In the present case, in particular, the following estimations have been found: $L(0.08$ Hz$) \approx 500$ km, $L(0.15$ Hz$) \approx 200$ km, and $L(0.24$ Hz$) \approx 100$ km.

As was mentioned above, the quantitative results obtained in this test cannot be compared to observations because of the absence of the latter. To a greater extent, this refers to the results obtained with one- and two-dimensional spectra (examples of evolution which were presented in paper [9] and monograph [1]). Nevertheless, in a physical aspect, the above qualitative results of items (1)–(3) are highly interesting and reliable. It also seems that the criterion mentioned in item (4) is very informative in comparing dissipation features of different models. Such a study is in progress at the present time for the WAM and WW models. However, analysis of these results will evidently need a special presentation.

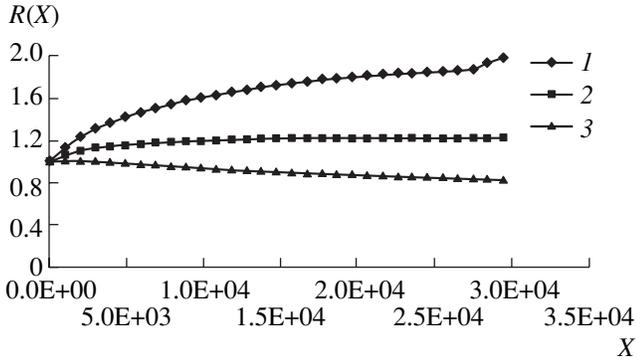

**Fig. 9.** Dependence of the ratio $R(X)$ on the dimensionless fetch $X = xg\,U_{10}^{-2}$ according to test 3. Parameters of the spectrum (51): (*1*) $R1 = 0.1$, $R2 = 2$; (*2*) $R1 = 0.1$, $R2 = 4$; and (*3*) $R1 = 0.1$, $R2 = 8$.

### 6.4. Swell along the Wind

**Formulation of the problem.** The wind is given as in the "straight fetch" test: $\mathbf{U}_{10}$ = const. The initial wave state is given by a homogeneous wave field for the spectrum of form (51), (52) with the following specifications. The wind-sea mode $S_1$ (taken with the coefficient $R1$) is given in the form described earlier in Section 6.2. The swell mode $S_2$ (taken with the coefficient $R2$) is defined by a special spectrum of the form (for explanation, see Subsection 6.3)

$$S_2(f, \theta, f_{psw}, \theta_{psw}, n, s_{sw})$$
$$= 0.1 \times 2\pi\sigma^{-n}\exp\left[-\frac{n}{4}(f_{0sw}/f)^4\right]\cos^{s_{sw}}(\theta - \theta_{psw}). \quad (56)$$

In (56), the integer value of the parameter $n$ is taken in the interval 6–10, the angular parameter $s_{sw}$ is chosen in the interval 4–12, and $\theta_{psw} = 0$. The initial peak frequency of swell is given by the condition $f_{0sw} < g/2\pi U_{10}$, while the parameters for the initial wind-sea spectrum are as follows: $f_0 \approx (1.5$–$2)g/2\pi U_{10}$, and $s = 2$. In particular, in the cases considered below, they were $f_{0sw} = 0.11$ Hz and $f_{0ws} \approx 0.24$ Hz, while the values for $R1$ and $R2$ were varied widely at $s_{sw} = 12$ and $n = 6$ and $10$. The boundary conditions correspond to those described in Subsection 6.1.

The aim of the test is to investigate the swell impact on the character of wind-sea development (swell–wind-sea-interaction test). Here are two questions of physical interest. The first question lies in determining the influence of developing waves on the character of swell decay. In such a case, the simplest control quantity is the relative energy in relation to the fetch $R(X)$ given by relation (54). The second question arises as to what extent the model can reproduce the effect of suppressing wind-sea development known from observations [41–43]. In such a case, as the control quantity, one may accept the value of the ratio of peak magnitudes of the spectrum fixed at the final stage of development for two options: in the presence of the swell $S_p^+$ and in the absence of swell $S_p^-$; i.e., the "suppression" parameter is given by

$$\nu \equiv S_p^+/S_p^-. \quad (57)$$

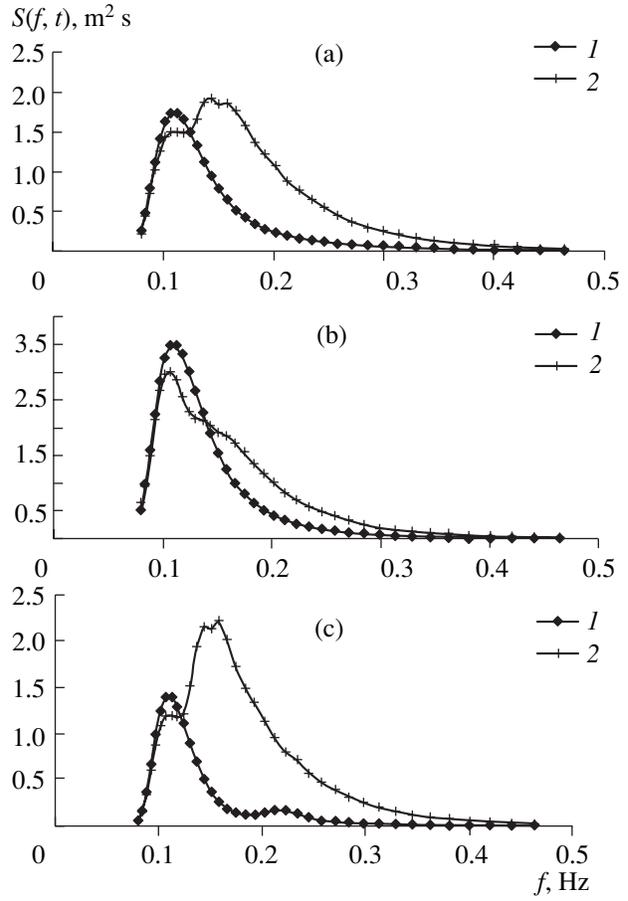

**Fig. 10.** Time history of the one-dimensional spectrum $S(f, t)$ according to test 3 at the boundary point of the calculation site for two dimensionless moments $T = tg/U_{10}$: (*1*) $T = 0$ and (*2*) $T = 10^5$ ($U_{10} = 10$ m/s). Parameters of spectra (51) and (56): (a) $R1 = 0.1$, $R2 = 2$ and $n = 6$, $s = 12$; (b) $R1 = 0.1$, $R2 = 4$ and $n = 6$, $s = 12$; and (c) $R1 = 0.1$, $R2 = 1$, and $n = 10$, $s = 12$.

The suppression parameter $\nu$ is analyzed at the final boundary point of the site $X_B = \Delta X N X$ in relation to the quantities $R1$ and $R2$, describing the composition of the initial wind field.

Results of these test calculations are shown in Figs. 9 and 10 for a wind with $U_{10} = 10$ m/s. Analysis of them leads to the following conclusions.

(i) In the presence of a swell, the development of a wind sea can lead to both the growth of wave energy (curves *1*, *2* in Fig. 9) and its decrease (curve *3*) in relation to the initial energy of a swell. The latter process occurs in the case when the initial energy of a swell exceeds the expected energy of a fully developed sea for the given wind. In any case, the total wave energy is 3–5% smaller than the energy of a developed sea. The cause of such reduction is related to the existence of two modes in the wave spectrum and is explained below.

(ii) The presence of the initial collinear swell leads to a marked reduction of the peak spectral value $S_p^+$ (at the stage of wind-sea establishment) with respect to that in the absence of swell $S_p^-$ (see Figs. 10a–10c and Fig. 7, respectively). This effect corresponds qualitatively to the observations mentioned in [41–43]. As follows from analysis of calculations for different values of the coefficients $R1$ and $R2$, this effect is provided by intensive nonlinear transfer of energy from the developing wave components to the swell components, which, in turn, lose the energy transferred to them in the course of natural dissipation. The existence of two modes in the spectrum also explains a rather small reduction of the total energy of the fully developed sea in the presence of a swell, because the swell mode partially compensates for a lack of energy of wind waves.

(iii) Comparison of curves in Figs. 10a and 10b with the curve in Fig. 7 gives an estimate for the suppression parameter of the order of $\nu \approx 0.65$–$0.7$. When the energy of a swell is greater than the energy of a fully developed wind sea, this effect is no longer dependent on the ratio of the coefficients $R1$ and $R2$. However, if the energy of a swell is rather small (a "weak swell"), $\nu$ approaches 1 with decreasing $R2$ (Fig. 10c).

The mechanism of the effect of suppressing a wind sea by a swell is explained by the role of nonlinear interactions among wind-wave components and a swell. (Details of its discussion can be found in [1, 9, 39].) It is of interest that, for the model proposed here, the effect of suppression is not as great as it is in the observation by Donelan [41] or in the model of [9]. The difference of the present results from those for the model of [9] is explained by a significant modification of the parametrization used in the new model for the Nl term. This example testifies that comparison of the results obtained here with analogous results obtained for other models is of remarkable scientific and methodological interest.

To conclude the section devoted to testing the wave part of the model, one may note that, in the future, this test may be elaborated as a separate investigation devoted to a better understanding of the physics of the effect under discussion.

### 6.5. Remarks on Testing the Dynamic Boundary Block

The methods and results of testing the features of the dynamic boundary block are described in detail in the recent paper [11]. From analysis of these results, it follows unambiguously that a significant variability of such parameters as the drag coefficient $C_d$ and Charnock's parameter $\tilde{z}_0 \equiv gz_0/u_*^2$ in relation to the wind or wave age (Figs. 2, 3) can be treated in a unified manner with the help of the dynamic boundary-layer model by Makin and Kudryavtsev [14]. Thus, we have confirmed both the validity of the functional rather than parametric approach to determining the aforementioned dependences and the impossibility of a simple recalculation of $U_{10}$ to $u_*$. Recommendations dealing with this issue are given in Section 3.

### 7. CONCLUSIONS

On the basis of analysis of numerous earlier results obtained by me, the parametrization of the source function of a numerical wind-wave model has been substantiated theoretically. This source function is optimized by the criterion of accuracy and speed of calculations. The optimized source function includes the following terms:

(a) a fast version of the discrete interaction approximation (FDIA) with a new configuration of four interacting wave vectors for the parametrization of the nonlinear mechanism of evolution (Section 2);

(b) a generalized empirical function of wave-energy input by wind improved by adding a special boundary-layer block (Section 3); and

(c) the wave-energy-dissipation term, which is squared in the wave spectrum and constructed on the basis of a semiphenomenological hydrodynamic theory taking into account the interaction of wave motion with the turbulence of the upper layer of water (Section 4).

The recalculation of the wind at the standard horizon $U_{10}$ to the friction velocity $u_*$ was considered specially. It was shown that such a recalculation can be considered justified only if one uses the dynamic boundary layer block based on a model similar to that proposed in [14] (Section 3). Including this block in a numerical wind-wave model imparts a new quality of a model of the next; that is, fourth, generation.

The wave (physical) part of the model was tested with the system of three tests, which are known as informative tests [1]: (1) straight fetch, (2) swell decay, (3) swell along the wind (Sections 5, 6). On the basis of a specially substantiated choice of control quantities, it was shown that the proposed model describes empirical dependences well (for the straight fetch) not only for the integral characteristics of waves but also for a number of parameters for the one-dimensional wave spectrum. Evidence was obtained that the system of tests used permits a better understanding of the physics of a series of effects observed during the evolution of wind-wave spectra.

It is expected that the results obtained will serve as a basis for comparison of the new model with the widely used models of the third generation WAM [4] and WW [6]. As a preliminary result of such a comparison, the fact is presented that the only substitution of the proposed parametrization of Nl in the FDIA version for the term of nonlinear evolution mechanism in the WAM model leads to a gain in the total time of wave forecasting of about 30% without any loss of accuracy [13]. The next stage of study will be the execution of a full-scale comparison of the models mentioned with the system of tests accepted here and their verification against full-scale data. After that, the optimized model will be elaborated to the level necessary for practical applications.

## ACKNOWLEDGMENTS

TI am grateful to G.S. Golitsyn and Yu.A. Volkov for the organizing support of the work and for numerous remarks during discussions. I thank the members of the Scientific Council of the Institute of Atmospheric Physics and the audience of V.P. Dymnikov's seminar at the Institute of Numerical Mathematics for interest in my work and constructive proposals.

This study was initiated at the Centre of Climate Study and Weather Forecasting in Brazil (CPTEC), where the author was a guest scientist, and supported by the Russian Foundation for Basic Research, project no. 04-05-64650.

610 POLNIKOV30. G. Rodrigues and C. G. Soares, "Uncertainty in the Estimation of the Slope of the High Frequency Tail of Wave Spectra," Appl. Ocean Res. **21**, 207–213 (1999).
31. M. L. Banner and X. Tian, "On the Determination of the Onset of Breaking for Modulating Surface Gravity Waves," J. Fluid Mech. **386**, 107–137 (1998).
32. M. A. Donelan, "Air–Water Exchange Processes (Physical Processes in Lakes and Oceans)," Coast. Estuar. Stud. **54**, 19–36 (1998).
33. A. V. Babanin, I. R. Young, and M. L. Banner, "Breaking Probabilities for Dominant Surface Waves on Water of Finite Constant Depth," J. Geophys. Res. C **106**, 11659–11676 (2001).
34. K. Hasselmann, "On the Spectral Dissipation of Ocean Waves due to White Capping," Boundary Layer Meteorol. **6**, 107–127 (1974).
35. M. L. Banner and I. R. Young, "Modeling Spectral Dissipation in the Evolution of Wind Waves. Part I. Assessment of Existing Model Performance," J. Phys. Oceanogr. **24**, 1550–1571 (1994).
36. V. K. Makin and M. Stam, *New Drag Formulation in NEDWAM. Technical Report no. 250* (KNMI, Netherlands, 2003).
37. V. V. Efimov and V. G. Polnikov, "Numerical Experiments on the Basis of a Model for Wind Waves with Turbulent Dissipation," Morsk. Gidrofiz. Zh., No. 1, 17–25 (1986).
38. V. V. Efimov and V. G. Polnikov, "Numerical Experiments on Modeling Wind Waves," Okeanologiya **25**, 725–732 (1985).
39. V. G. Polnikov, "On Description of a Wind-Wave Energy Dissipation Function," in *Proceedings of Air–Sea Interface Symposium, Marseilles, France* (Marseilles Univ., Marseilles, 1994), pp. 227–282.
40. V. G. Polnikov, *Study of Nonlinear Interactions in the Spectrum of Wind Waves*, Doctoral Dissertation in Mathematics and Physics (MGI NANU, Sevastopol, 1995).
41. M. A. Donelan, "The Effect of Swell on the Growth of Wind Waves," Johns Hopkins Univ. Tech. Digest **8** (1), 18–23 (1987).
42. H. Mitsuyasu and Y. Yoshida, "Air–Sea Interactions under the Existence of Swell Propagating against the Wind," Bull. Res. Inst. Appl. Mech. Kyushi Univ. **63**, 47–71 (1989).
43. N. Booij, L. H. Holthuijsen, and I. J. Haagsma, *The Effect of Swell on the Generation and Dissipation of Wind Sea. Ocean Wave Measurement and Analysis*, Ed. by B. Edge and I. M. Hemsley (ASCE, San Francisco, 2001), pp. 501–506.
IZVESTIYA, ATMOSPHERIC AND OCEANIC PHYSICS    Vol. 41    No. 5    2005